\documentclass[12pt]{article}
\usepackage{graphicx}
\usepackage{bm}

\newcommand\pubnumber{}
\newcommand\pubdate{}

\def\kias{School of Physics, KIAS,
Seoul 130-722, Korea}
\def\ku{Department of Physics, Korea University,
Seoul 136-713, Korea}

\def\Title#1{\begin{center} {\Large #1 } \end{center}}
\def\Author#1{\begin{center}{ \sc #1} \end{center}}
\def\Address#1{\begin{center}{ \it #1} \end{center}}

\newcommand\pubblock{\rightline{\begin{tabular}{l} \pubnumber\\
         \pubdate  \end{tabular}}}
\newenvironment{Abstract}{\begin{quotation}  }{\end{quotation}}
\newenvironment{Presented}{\begin{quotation} \begin{center} 
             PRESENTED AT\end{center}\bigskip 
      \begin{center}\begin{large}}{\end{large}\end{center} \end{quotation}}
\def\Acknowledgements{\bigskip  \bigskip \begin{center} \begin{large}
             \bf ACKNOWLEDGEMENTS \end{large}\end{center}}

\def\beq{\begin{equation}}
\def\eeq#1{\label{#1}\end{equation}}
\def\eeqn{\end{equation}}

\def\beqa{\begin{eqnarray}}
\def\eeqa#1{\label{#1}\end{eqnarray}}
\def\eeqan{\end{eqnarray}}

\let\bar=\overbar

\def\Dslash{\not{\hbox{\kern-4pt $D$}}}
\def\dslash{\not{\hbox{\kern-2pt $\del$}}}

\def\msb{{\bar{\ssstyle M \kern -1pt S}}}

\begin{document}
\begin{titlepage}
\pubblock

\vfill
\Title{Higher-order corrections to exclusive production 
\\of charmonia at $B$ factories}
\vfill
\Author{ Ying Fan and Jungil Lee}
\Address{\ku}
\Author{ Chaehyun Yu}
\Address{\kias}
\vfill
\begin{Abstract}
As a test of the color-singlet mechanism of the nonrelativistic QCD 
(NRQCD) factorization approach, we consider  
the exclusive two-quarkonium productions in electron-positron annihilation
$e^+ e^- \to \eta_c + \gamma$ and $e^+ e^- \to J/\psi + J/\psi$ 
at $B$ factories.
The cross sections are computed to the next-to-leading order  
in $\alpha_s$ and are resummed to all orders in half the relative 
velocity $v$ of the charm quark in each meson rest frame.
The available theoretical prediction of the cross section for 
$e^+ e^- \to J/\psi + \eta_c$ at the same level of theoretical accuracies
is consistent with the available experimental data.
Those for $e^+ e^- \to \eta_c + \gamma$ and 
$e^+ e^- \to J/\psi + J/\psi$ that are computed new in this work
can be tested against the data from future super $B$ factories.
\end{Abstract}
\vfill
\begin{Presented}
The 5th International Workshop on Charm Physics\\
Honolulu, USA,  May 14--17, 2012
\end{Presented}
\vfill
\end{titlepage}
\def\thefootnote{\fnsymbol{footnote}}
\setcounter{footnote}{0}

\section{Introduction}
Production of heavy quarkonium provides a unique opportunity to probe 
the framework of the nonrelativistic quantum chromodynamics 
(NRQCD) factorization approach \cite{NRQCD}, 
which is an effective field theory to describe production and decay 
of heavy quarkonium. In this approach, a production or decay rate 
of a heavy quarkonium is expressed as a linear combination
of NRQCD long-distance matrix elements (LDMEs) and the short-distance 
coefficients of each LDMEs are insensitive to the long-distance nature
of hadrons. Each LDME represents the transition rate of a heavy quark
and antiquark ($Q\bar{Q}$) pair with a specific spectroscopic state
to evolve into the physical quarkonium state. The NRQCD factorization
for the decay is proved in Ref. \cite{NRQCD}. While proofs of 
factorization for some of exclusive quarkonium production processes
are available \cite{fac}, the factorization in the inclusive quarkonium
production is still a conjecture.

In order to predict the production cross section of a heavy quarkonium,
it is required to determine all of the relevant NRQCD LDMEs.
The universality of the LDMEs requires that the LDMEs are independent
of specific processes. In addition, an LDME for a production is the same
as that for the decay under the vacuum-saturation approximation of order 
$v^4$, where $v$ is half the relative velocity of the heavy quark in the
quarkonium rest frame \cite{NRQCD}. At present, the LDMEs are determined
by comparing the NRQCD factorization formulas with the measured production
and decay rates of a heavy quarkonium.
The color-singlet LDMEs can be determined from the electromagnetic decays
of a heavy quarkonium \cite{Bodwin:2007fz,Chung:2010vz}.
In the case of the color-octet channels, they use the inclusive production
rates of a heavy quarkonium at various colliders, Tevatron, LHC, HERA, 
$B$ factories, and so on \cite{co}. Because the color-singlet channel
also contributes to the inclusive production processes, the color-singlet
LDMEs that are determined form the electromagnetic decays are used as
input parameters.

Let us focus on the color-singlet LDMEs for the $S$-wave
heavy quarkonia, in particular, $J/\psi$ and $\eta_c$ that depend
dominantly only the color-singlet channel. The corresponding LDMEs are
determined from $J/\psi \to \ell^+ \ell^-$ and 
$\eta_c\to \gamma\gamma$ \cite{Bodwin:2007fz} and applied to various 
production processes involving $J/\psi$ or $\eta_c$.
The cross section for $e^+ e^- \to J/\psi + \eta_c$ at the $B$ factories 
has substantial contributions 
from the relativistic \cite{Bodwin:2007ga,He:2007te} as well
as QCD corrections \cite{Zhang:2005ch}. 
The prediction \cite{Bodwin:2007ga,He:2007te} 
that includes both relativistic and QCD corrections is consistent with
the data \cite{Abe:2002rb,Abe:2004ww} within errors.
Therefore, both relativistic and QCD corrections
may have significant contributions in other exclusive $S$-wave quarkonium 
production in electron-positron annihilation.
In this work, we compute the relativistic corrections, 
resummed to all orders in $v$,
to the cross sections for the exclusive processes
$e^+ e^- \to \eta_c + \gamma$ and
$e^+ e^- \to J/\psi + J/\psi$ that have not been observed at the $B$
factories, yet. Then, we try to combine them together with the QCD 
corrections in order to present a more reliable prediction for the cross
section.
The theoretical prediction can be tested against the data
from future experiments at BELLE II or super $B$ factory.

\section{Higher-order corrections}
A schematic form of the differential cross section 
of a heavy quarkonium is 
$d \sigma \sim d \hat{\sigma}_n \langle O_n^H \rangle$,
where $\langle O_n^{H} \rangle$ is the NRQCD LDME representing the 
transition of a $Q\bar{Q}$ pair with a spectroscopic state $n$
($Q\bar{Q}_n$)
to a heavy quarkonium, $d \hat{\sigma}_n$ is the corresponding 
short-distance coefficient involving the production of the 
pair $Q\bar{Q}_n$ from the initial state, and the summation over $n$
is implicit. Because we consider only the color-singlet channels
in this work, we use the corresponding LDME $\langle O_n\rangle_H$, 
that is accurately determined from the electromagnetic decay,
for the decay by applying the vacuum-saturation approximation
$\langle O_n^H\rangle\approx (2J+1)\langle O_n\rangle_H$, where 
$J$ is the total-angular-momentum quantum number of $H$.

The short-distance coefficient $d \hat{\sigma}_n$ is 
a perturbative series in
the strong coupling $\alpha_s$. 
In various quarkonium production processes, $d\hat{\sigma}_n$ are 
known to the next-to-leading-order (NLO) accuracies in $\alpha_s$.
The $K$ factor in each process depends on the choice of input variables
like the heavy-quark mass $m_Q$ and the factorization scale $\mu$.
Complete next-to-next-to-leading-order (NNLO) corrections in $\alpha_s$ 
are known only in electromagnetic decays of a heavy quarkonium,
$J/\psi \to \ell^+ \ell^-$ and $\eta_c \to \gamma \gamma$ \cite{Beneke:1997jm}
from which one can determine the color-singlet LDMEs.
However, unless one computes the short-distance coefficients for another
specific process to the same accuracies (NNLO in $\alpha_s$), the prediction 
may depend strongly on the factorization scale. 
Because the short-distance coefficients for
$e^+e^-\to J/\psi+\eta_c$,
$e^+e^-\to \eta_c+\gamma$, and
$e^+e^-\to J/\psi+J/\psi$, that we consider in this work, 
have not been computed to NNLO in $\alpha_s$,
we use the numerical values for the
color-singlet NRQCD LDMEs that were determined from the
electromagnetic decays at order $\alpha_s$.

At leading order (LO) in $\alpha_s$
the spin-triplet $S$-wave heavy-quarkonium production 
consists of QCD and QED processes.
Among QED diagrams, the photon-fragmentation process in which a virtual
photon fragments into the color-singlet spin-triplet $S$-wave 
$Q\bar{Q}$ pair, can be comparable to the QCD process.
This enhancement happens when $M^*\gg m_H$,
where $M^*$ is the typical virtuality of the internal lines
other than the fragmenting photon and the quarkonium mass
$m_H$ represents the virtuality of the fragmenting photon. 
In $e^+e^-$ annihilation
$M^*\sim $ half the center-of-momentum energy ($\sqrt{s}$) and
in hadron collisions
$M^*\sim$ the transverse momentum $p_T$ of the quarkonium.
In these limits, the enhancement due to the propagator denominator
of the fragmenting photon overcomes the strong suppression
factor $(\alpha/\alpha_s)^2$ of the QED process relative to the QCD process.
In the $J/\psi$ production at hadron colliders, the QED
contribution via photon fragmentation
could be larger than that from the usual QCD process at sufficiently large
$p_T$ \cite{He:2009cq}. In $e^+ e^- \to J/\psi +\eta_c$ 
at $B$ factories, the QED contribution through photon fragmentation can reach
about 19\,\% of the QCD contribution at LO in 
$\alpha_s$ \cite{Bodwin:2007ga}.

In addition, the order $\alpha_s v^2$ correction is also
a potential source of large corrections. The corrections are available
for $J/\psi \to \ell^+ \ell^-$ \cite{Bodwin:2008vp},
$B_c \to \ell \nu$ \cite{Lee:2010ts}, 
$\eta_c \to \gamma \gamma$ and light hadrons \cite{Jia:2011ah},
and $e^+ e^- \to J/\psi + \eta_c$ \cite{Dong:2012xx}.
In $J/\psi \to \ell^+ \ell^-$
and $B_c \to \ell \nu$, the relativistic corrections at order $\alpha_s$
are resummed to all orders in $v$.
In $S$-wave heavy quarkonium decays,
the order $\alpha_s v^2$ corrections are not so large. For example, 
in the $J/\psi \to \ell^+ \ell^-$ decay, the relativistic corrections
at order $\alpha_s$ are at most $0.3$\,\% \cite{Bodwin:2008vp}.
In the case of $e^+ e^- \to J/\psi + \eta_c$ at $B$ factories, 
the relativistic corrections at order $\alpha_s$ 
enhance the cross section mildly \cite{Dong:2012xx}.

In the heavy-quarkonium process, both the QCD 
and relativistic corrections could be large. In this case, the
interference between the amplitude at NLO in $\alpha_s$ and 
that for the relativistic corrections
might also be large. For example, in the $e^+ e^- \to J/\psi + \eta_c$ process,
the $K$ factor from the QCD corrections is about 2 and the relativistic 
corrections to the short-distance coefficients increase the cross section
by a factor of 40\,\% \cite{Bodwin:2007ga}. Then, the interference
can reach about 26\,\% of the QCD process at LO in $\alpha_s$.

\section{Relativistic corrections to an $\bm{S}$-wave quarkonium}
The order-$v^{2n}$ correction to $S$-wave quarkonium ($H$) process
in the color-singlet channel
is proportional to the ratio $\langle \bm{q}^{2n} \rangle_H$
of the LDME of relative order $v^{2n}$ to the LO one 
$\langle O_1 \rangle_H$ \cite{Bodwin:2007fz}.
Here, $\bm{q}$ is the spatial component of half the relative momentum 
of the $Q$ and $\bar{Q}$ in the $Q\bar{Q}$ rest frame.
According to Ref.~\cite{Bodwin:2006dm} the amplitude $A[H]$ 
expanded to all orders in $v$ can be computed from
the hard amplitude $T(\bm{q}^2)$ as 
\begin{equation}
A[H] = \sum_n \left. \left[
\frac{1}{n!}
\left( \frac{\partial}{\partial \bm{q}^2} \right)^n
T(\bm{q}^2) \right] \right|_{\bm{q}^2=0}
\langle \bm{q}^{2n}\rangle_H 
\langle O_1 \rangle_H^{1/2}.
\label{expan}%
\end{equation}
If we make use of the relation
$\langle \bm{q}^{2n}\rangle_H\approx \langle \bm{q}^{2}
\rangle_H^n$ \cite{Bodwin:2006dm},
a generalized version of the Gremm-Kapustin relation
\cite{Gremm:1997dq},
we can resum a class of relativistic corrections
to all orders in $v$. Then, the expansion (\ref{expan}) is simplified as
\begin{equation}
\label{eq:A-resummed}
A[H] = T\left( \langle \bm{q}^2 \rangle_H \right)
\langle O_1 \rangle_H^{1/2}.
\end{equation}
Because the relation 
$\langle \bm{q}^{2n}\rangle_H\approx \langle \bm{q}^{2}
\rangle_H^n$ \cite{Bodwin:2006dm}
has errors of relative order $v^2$ due to the neglect of 
the spin-dependent potential and
the gauge
field contribution to the covariant derivative in the LDME in the 
Coulomb gauge \cite{Bodwin:2006dm}, Eq.~(\ref{eq:A-resummed}) is accurate 
to order $v^4$.
However, the resummation of a class of relativistic 
corrections may still be useful in estimating the size of the complete 
relativistic corrections. If the resummed factorization formula involves
uncomfortably large corrections, then one may cast doubt on the convergence
of the series. Once the resummed formula gives moderate corrections, then
one may treat this as a clue that the series may converge well.
Another strong point of this resummation method is that the computation
is far easier than the fixed-order relativistic corrections that involve
a large number of terms generated by the derivatives. The method is 
particularly effective in $e^+ e^-\to J/\psi+J/\psi$ that we present later.

\section{$\bm{e^+ e^- \to J/\psi + \eta_c}$}
In this section, we consider the process $e^+ e^- \to J/\psi + \eta_c$
at $B$ factories. This process proceeds through one photon exchange
between leptonic and hadronic currents.
In the early stage,
there was a discrepancy by an order of magnitude between 
theoretical predictions and experiments 
for the cross section \cite{Abe:2002rb,Braaten:2002fi}.
Later, the discrepancy was resolved due to the improvements
in both theory and experiment.
The measured cross section has decreased compared to the first measurement
\cite{Abe:2004ww} and 
the corresponding theoretical predictions were enhanced after including
order-$\alpha_s$ and relativistic 
corrections \cite{Bodwin:2007ga,He:2007te,Zhang:2005ch}.
The $K$ factor from NLO corrections in $\alpha_s$ is about 2, which depends
on the charm-quark mass $m_c$ \cite{Zhang:2005ch}.
The pure relativistic corrections consist of direct and indirect
contributions that come from the corrections to the
short-distance coefficient and through the LDME
that amount to 40\,\% and 72\,\% of LO contribution, 
respectively \cite{Bodwin:2007ga}.
The interference between the QCD and relativistic corrections
was also computed in Ref.~\cite{Bodwin:2007ga}. Then, the total cross section 
is $17.6^{+7.8}_{-6.3}$ fb \cite{Bodwin:2007ga}, which is consistent with 
the empirical data
$25.6\pm 4.4$ fb at Belle and $17.6\pm 3.5$ fb at BABAR \cite{Abe:2004ww}.
Recently, improvement to this process at $O(\alpha_s v^2)$ has been
carried out, but the cross section is enhanced mildly \cite{Dong:2012xx}.

The process $e^+ e^- \to J/\psi + \eta_c$ has been proved to have a 
significant relativistic corrections in comparison with any other
quarkonium process. Now we do not have serious discrepancy between theory
and experiment regarding this process. However, there still remain some
subtle issues. The current experimental data for
$e^+ e^- \to J/\psi + \eta_c$
come from $\mu^+\mu^-+$ at least two charged tracks,
which account for the $\eta_c$ decay. If one further includes the
events without charged tracks, then the measured cross section 
can be bigger than the current data.
It is interesting to see if the uncalculated corrections at NNLO 
in $\alpha_s$ enhances the theoretical prediction to catch up with
the possible experimental enhancement.
If it is not the case, then the discrepancy between the theory
and experiment may revive.

\section{$\bm{e^+ e^- \to \eta_c + \gamma}$}
In this section, we consider the process $e^+ e^- \to \eta_c + \gamma$.
This process proceeds through one photon exchange because the 
charge-conjugation parity in the final state is $-1$ \cite{Chung:2008km}. 
Similarly, one may also consider
the production of any heavy quarkonium with charge conjugation parity $+1$
associated with a photon.
This process was suggested to be a good probe
to the color-singlet mechanism of NRQCD, 
especially for $\eta_c(2S)$ \cite{Chung:2008km}. Later, the NLO corrections
in $\alpha_s$ and the relativistic corrections at order $v^2$ were
computed \cite{Sang:2009jc}.

At LO in $\alpha_s$ and $v$, the cross section for $e^+ e^- \to \eta_c +
\gamma$ at $\sqrt{s}=10.58$ GeV is $82^{+21.4}_{-19.8}$ fb 
for $\langle O_1 \rangle_{\eta_c}=0.437^{+0.111}_{-0.105}$ 
GeV$^3$ \cite{Chung:2008km}. The NLO corrections in $\alpha_s$ decrease
the cross section by 18\,\% and the relativistic corrections at order $v^2$
reduce the cross section by 12\,\% for $v^2=0.13$ \cite{Sang:2009jc}.
However, this $v^2$ value is rather underestimated compared with
the conventional value $0.3$ for a charmonium. If one uses $v^2=0.23$
determined from the Cornell potential and resummed formula for 
the electromagnetic decay rates of $J/\psi$ and $\eta_c$ \cite{Bodwin:2007fz},
then the order-$v^2$ corrections can reach 21\,\%. We carry out the resummation
of relativistic corrections to all orders in $v$ and find that
the cross section decreases by 17\,\%, which is slightly smaller than 
the $v^2$ corrections \cite{progress}. This implies that the $v^2$ expansion
in this process converges rapidly. 
The relativistic corrections are comparable
to the NLO corrections in $\alpha_s$. 
This indicates that the inclusion of both 
QCD and relativistic corrections may improve the predictive power
of the theoretical prediction.
Furthermore, it might be necessary to compute interference between
the corrections of NLO in $\alpha_s$ and 
the relativistic corrections.
Taking into account all of the corrections
listed above, we find that
$\sigma[e^+e^-\to\eta_c+\gamma]=55.1$ fb for $\mu=2 m_c$ \cite{progress}.

\section{$\bm{e^+ e^- \to J/\psi + J/\psi}$}
In this section, we consider the process
 $e^+ e^- \to J/\psi + J/\psi$.
This process proceeds through two photon exchange because
both $J/\psi$ and $\gamma$ are in odd parities under charge conjugation.
This process was originally suggested to resolve the $e^+ e^- \to J/\psi+
\eta_c$ puzzle \cite{Bodwin:2002fk}. However, the Belle Collaboration found 
no evidence of this process and set an upper bound for the cross section
times the branching ratio of one of two $J/\psi$'s decaying 
into at least two charged
particles to be $9.1$ fb \cite{Abe:2004ww}. The angular distribution analysis
of $e^+ e^- \to J/\psi + \eta_c$ events by the Belle Collaboration
also disfavored the double $J/\psi$ production at $B$ factories.

There are four Feynman diagrams in this process. Two of them are photon 
fragmentation diagrams, where each virtual photon evolves into a $J/\psi$.
The remaining two are called the nonfragmentation diagrams.
The photon fragmentation diagrams dominate over the nonfragmentation
diagrams \cite{Bodwin:2006yd}. The nonfragmentation contribution
is at most 0.1\,\% of the fragmentation contribution.
The contributions of the interference between
the fragmentation and nonfragmentation diagrams is about
13\,\% of the fragmentation contribution \cite{Bodwin:2002fk}.
As is stated earlier, the dominance of the fragmentation contribution
is due to the large enhancement factor from the propagator denominator
of the fragmenting photon whose virtuality is of order $m_{J/\psi}$.
Because each contribution makes a separate gauge-invariant subset,
gauge invariance remains although we use different strategies
to compute each set of the amplitude.
In Ref. \cite{Bodwin:2006yd}, the authors employed
the vector-meson-dominance (VMD) method for the photon fragmentation
diagrams by replacing the photon-to-$J/\psi$ vertex by a coupling 
$g_{J/\psi}$, which is determined from the leptonic decay of $J/\psi$.
This method includes automatically the relativistic and
QCD corrections to the fragmentation contribution.
In the case of the nonfragmentation diagrams, they used the 
standard NRQCD approach to compute the amplitude.
Then, the total cross section for $e^+ e^- \to J/\psi + J/\psi$
at $B$ factories is $1.69\pm 0.35$ fb, which is well below
the upper bound at Belle \cite{Bodwin:2006yd}.

The computation of the NLO corrections in $\alpha_s$ was carried out 
within the framework of NRQCD \cite{Gong:2008ce}. The $K$ factor strongly 
depends on $m_c$ and the factorization scale $\mu$.
For $\mu=2 m_c$, the $K$ factor is about $0.077$ for $m_c=1.5$ GeV
and about $0.057$ for $m_c=1.4$ GeV, respectively \cite{Gong:2008ce}.
We note that most of the NLO corrections in $\alpha_s$ come from 
the corrections to the photon fragmentation diagrams. If the VMD treatment
is applied to the fragmentation diagrams, the $K$ factor can be large.

As we have mentioned, the resummation of relativistic corrections to all 
orders in $v$ is far easier than the computation of fixed-order
relativistic corrections. This is particularly true in this case 
because of the $t$-channel electron propagator that is an additional
source of relativistic corrections. We find that the resummation
of relativistic corrections decreases the cross section 
by 58\,\% \cite{progress}. 
The sum of the relativistic corrections and the corrections of NLO 
in $\alpha_s$ can be even negative because both corrections are
quite large. The inclusion of the
interference between the relativistic and QCD NLO corrections
may be helpful to cure the problem of the negative cross section
because the interference is positive.
Therefore, one must be very careful 
in combining these corrections, especially in the process
$e^+ e^- \to J/\psi + J/\psi$.

\section{Summary}
Exclusive heavy-quarkonium production in electron-positron collisions
provides a unique opportunity
to test the color-singlet mechanism of NRQCD. In this work,
we have discussed three exclusive processes, $e^+ e^- \to J/\psi + \eta_c$, 
$e^+ e^- \to \eta_c + \gamma$, and $e^+ e^- \to J/\psi + J/\psi$
at $B$ factories. 
The available theoretical prediction of the cross section for 
$e^+ e^- \to J/\psi + \eta_c$ is consistent with the experimental values
measured by the Belle and BABAR 
collaborations within uncertainties.
We have computed the relativistic corrections to the cross sections for
$e^+ e^- \to \eta_c + \gamma$ and $e^+ e^- \to J/\psi + J/\psi$
resummed to all orders in $v$. The predictions are further improved  by
adding additional corrections of NLO in $\alpha_s$. These new predictions
can be tested against the data from future super $B$ factories.


\Acknowledgements
This work is supported in part by Basic Science Research Program
through NRF 2011-0022996 and by NRF Research Grant
2012R1A2A1A01006053.


\begin{thebibliography}{99}


\bibitem{NRQCD} 
  G.~T.~Bodwin, E.~Braaten, and G.~P.~Lepage,
  Phys.\ Rev.\ D {\bf 51}, 1125 (1995)  
  [Erratum-ibid.\ D {\bf 55}, 5853 (1997)]  
  [hep-ph/9407339].  

\bibitem{fac} 
  G.~T.~Bodwin, X.~Garcia i Tormo, and J.~Lee,
  Phys.\ Rev.\ Lett.\  {\bf 101}, 102002 (2008)  [arXiv:0805.3876 [hep-ph]];
  Phys.\ Rev.\ D {\bf 81}, 114014 (2010)  [arXiv:1003.0061 [hep-ph]];  
  Phys.\ Rev.\ D {\bf 81}, 114005 (2010)  [arXiv:0903.0569 [hep-ph]].

\bibitem{Bodwin:2007fz} 
  G.~T.~Bodwin, H.~S.~Chung, D.~Kang, J.~Lee, and C.~Yu,
  Phys.\ Rev.\ D {\bf 77}, 094017 (2008)  [arXiv:0710.0994 [hep-ph]].  

\bibitem{Chung:2010vz} 
  H.~S.~Chung, J.~Lee, and C.~Yu,
  Phys.\ Lett.\ B {\bf 697}, 48 (2011)  [arXiv:1011.1554 [hep-ph]].  

\bibitem{co}
  M.~Butenschoen, and B.~A.~Kniehl,
  Phys.\ Rev.\ D {\bf 84}, 051501 (2011)  [arXiv:1105.0820 [hep-ph]];  
  Y.~-Q.~Ma, K.~Wang, and K.~-T.~Chao,
  Phys.\ Rev.\ D {\bf 84}, 114001 (2011)  [arXiv:1012.1030 [hep-ph]].  

\bibitem{Bodwin:2007ga} 
  G.~T.~Bodwin, J.~Lee, and C.~Yu,
  Phys.\ Rev.\ D {\bf 77}, 094018 (2008)  [arXiv:0710.0995 [hep-ph]];  
  G.~T.~Bodwin, D.~Kang, T.~Kim, J.~Lee, and C.~Yu,
  AIP Conf.\ Proc.\  {\bf 892}, 315 (2007)  [hep-ph/0611002].  

\bibitem{He:2007te} 
  Z.~-G.~He, Y.~Fan, and K.~-T.~Chao,
  Phys.\ Rev.\ D {\bf 75}, 074011 (2007)  [hep-ph/0702239 [HEP-PH]].  

\bibitem{Zhang:2005ch}
  Y.~J.~Zhang, Y.~j.~Gao, and K.~T.~Chao,
  Phys.\ Rev.\ Lett.\  {\bf 96}, 092001 (2006)
  [arXiv:hep-ph/0506076];
  B.~Gong and J.~-X.~Wang,
  Phys.\ Rev.\ D {\bf 77}, 054028 (2008)  [arXiv:0712.4220 [hep-ph]].  

\bibitem{Abe:2002rb} 
  K.~Abe {\it et al.}  [Belle Collaboration],
  Phys.\ Rev.\ Lett.\  {\bf 89}, 142001 (2002)  [hep-ex/0205104].  

\bibitem{Abe:2004ww} 
  K.~Abe {\it et al.}  [Belle Collaboration],
  Phys.\ Rev.\ D {\bf 70}, 071102 (2004)  [hep-ex/0407009];  
  B.~Aubert {\it et al.}  [BABAR Collaboration],
  Phys.\ Rev.\ D {\bf 72}, 031101 (2005)  [hep-ex/0506062].  

\bibitem{Beneke:1997jm} 
  M.~Beneke, A.~Signer, and V.~A.~Smirnov,
  Phys.\ Rev.\ Lett.\  {\bf 80}, 2535 (1998)  [hep-ph/9712302];  
  A.~Czarnecki and K.~Melnikov,
  Phys.\ Lett.\ B {\bf 519}, 212 (2001)  [hep-ph/0109054].  

\bibitem{He:2009cq} 
  Z.~-G.~He, R.~Li, and J.~-X.~Wang,
  Phys.\ Lett.\ B {\bf 711}, 371 (2012)  [arXiv:0904.1477 [hep-ph]].  

\bibitem{Bodwin:2008vp} 
  G.~T.~Bodwin, H.~S.~Chung, J.~Lee, and C.~Yu,
  Phys.\ Rev.\ D {\bf 79}, 014007 (2009)  [arXiv:0807.2634 [hep-ph]].  

\bibitem{Lee:2010ts} 
  J.~Lee, W.~Sang, and S.~Kim,
  JHEP {\bf 1101}, 113 (2011)  [arXiv:1011.2274 [hep-ph]].  

\bibitem{Jia:2011ah} 
  Y.~Jia, X.~-T.~Yang, W.~-L.~Sang, and J.~Xu,
  JHEP {\bf 1106}, 097 (2011)  [arXiv:1104.1418 [hep-ph]];  
  H.~-K.~Guo, Y.~-Q.~Ma, and K.~-T.~Chao,
  Phys.\ Rev.\ D {\bf 83}, 114038 (2011)  [arXiv:1104.3138 [hep-ph]].  

\bibitem{Dong:2012xx} 
  H.~-R.~Dong, F.~Feng, and Y.~Jia,
  Phys.\ Rev.\ D {\bf 85}, 114018 (2012)  [arXiv:1204.4128 [hep-ph]].  

\bibitem{Bodwin:2006dm} 
  G.~T.~Bodwin, D.~Kang, and J.~Lee,
  Phys.\ Rev.\ D {\bf 74}, 114028 (2006)  [hep-ph/0603185];  
  Phys.\ Rev.\ D {\bf 74}, 014014 (2006)  [hep-ph/0603186].  

\bibitem{Gremm:1997dq} 
  M.~Gremm and A.~Kapustin,
  Phys.\ Lett.\ B {\bf 407}, 323 (1997)  [hep-ph/9701353].  

\bibitem{Braaten:2002fi}
  E.~Braaten and J.~Lee,
  Phys.\ Rev.\  D {\bf 67}, 054007 (2003)
  [Erratum-ibid.\  D {\bf 72}, 099901 (2005)]
  [arXiv:hep-ph/0211085];
  K.~Y.~Liu, Z.~G.~He, and K.~T.~Chao,
  Phys.\ Lett.\  B {\bf 557}, 45 (2003)
  [arXiv:hep-ph/0211181].

\bibitem{Chung:2008km} 
  H.~S.~Chung, J.~Lee, and C.~Yu,
  Phys.\ Rev.\ D {\bf 78}, 074022 (2008)  [arXiv:0808.1625 [hep-ph]].  

\bibitem{Sang:2009jc} 
  W.~-L.~Sang and Y.~-Q.~Chen,
  Phys.\ Rev.\ D {\bf 81}, 034028 (2010)  [arXiv:0910.4071 [hep-ph]];  
  D.~Li, Z.~-G.~He, and K.~-T.~Chao,
  Phys.\ Rev.\ D {\bf 80}, 114014 (2009)  [arXiv:0910.4155 [hep-ph]].  

\bibitem{progress}
  Y.~Fan, J.~Lee, and C.~Yu, work in progress.

\bibitem{Bodwin:2002fk} 
  G.~T.~Bodwin, J.~Lee, and E.~Braaten,
  Phys.\ Rev.\ Lett.\  {\bf 90}, 162001 (2003)  [hep-ph/0212181];  
  Phys.\ Rev.\ D {\bf 67}, 054023 (2003)  
  [Erratum-ibid.\ D {\bf 72}, 099904 (2005)]  [hep-ph/0212352].  

\bibitem{Bodwin:2006yd} 
  G.~T.~Bodwin, E.~Braaten, J.~Lee, and C.~Yu,
  Phys.\ Rev.\ D {\bf 74}, 074014 (2006)  [hep-ph/0608200].  

\bibitem{Gong:2008ce} 
  B.~Gong and J.~-X.~Wang,
  Phys.\ Rev.\ Lett.\  {\bf 100}, 181803 (2008)  [arXiv:0801.0648 [hep-ph]].  


\end{thebibliography}
\end{document}